# Microgroove convexity is critical for robust gaseous layers on hierarchically-structured superhydrophobic surfaces

Xiao Han[1], Jingnan Liu[1], Moneesh Upmanyu[2]*, Hailong Wang[1]*

[1]CAS Key Laboratory of Mechanical Behavior and Design of Materials, Department of Modern Mechanics, University of Science and Technology of China, Hefei, Anhui 230027, China
[2]Group for Simulation and Theory of Atomic-Scale Material Phenomena (stAMP), Department of Mechanical and Industrial Engineering, Northeastern University, Boston, Massachusetts 02115, USA
* Corresponding author. Email: mupmanyu@neu.edu (M.U.); hailwang@ustc.edu.cn (H.W.)

**Abstract**

Gaseous layers (plastrons) trapped on the surfaces of immersed hydrophobic surfaces are critical for their function. Fibrillar morphologies offer a natural pathway, yet they are limited to a narrow range of liquid-surface systems and are vulnerable to pressure fluctuations that irreversibly destroy the plastron. Inspired by the convexly grooved surfaces of water fern (*Salvinia*) leaves that support their fibrous outgrowths, we study the plastron formation on 3D-printed dual-scale surfaces with elliptical interconnected microgrooves. The groove curvature stabilizes a seed gas layer (SGL) that facilitates plastron formation and restoration for all immersed hydrophobic surfaces. Computations and theoretical calculations reveal that the SGL storage capacity that sets the plastron robustness follows from the liquid menisci adaption to the groove geometry and pressure, and it can be further tuned using separated grooves. Our study highlights groove convexity as a key morphological feature for the design of multi-scale immersed surfaces for robust superhydrophobicity.

**Introduction**

Underwater superhydrophobic surfaces (SHPOSs) capped by a gaseous layer dramatically improve drag reduction[1-2] and marine antifouling[3], enabling a diverse set of applications for aquatic locomotion, liquid-repellency[4-6], self-cleaning[7-8], anti-icing[9-11]. Promising SHPOS architectures based on the super-repellent surfaces of natural organisms such as *Salvinia* leaves[12-14] and springtails[15-18] have shown strong repellency to liquids with ultra-low surface tension. However, the effectiveness of these microstructures has been limited to high density engineered filaments, primarily due to instabilities induced by the liquid pressure and its fluctuations that lead to a loss of the gaseous plastron, or the Wenzel state. Hierarchically structured topographies[12, 19-20] composed of a (first-level) micro-scale fibrillar coating supported by (second-level) surface nano-scale fibrillar coatings[19-20] or microgrooves[12] (Fig. 1a) are effective in seed gas layer (SGL) trapping within the second-level structures, aiding the formation and restoration of the plastron between the (first-level) fibers, or the Cassie state.

Reversing the Cassie-Baxter-Wenzel transition can be otherwise challenging and often prohibitively expensive as it requires external stimuli such as mechanical vibration[21], electrical generation[22], direct bubble replenishment[23-24], heating[25-26], and chemical catalysis[27]. The effect of the (second-level) grooved microstructure has garnered much less attention. In a recent study motivated by grooved epidermal surfaces of *Salvinia* leaves, a V-shaped wedged morphology was found to trap a gas layer and thereby stabilize the gaseous plastron. Close examination of the surfaces of the *Salvinia* leaves reveal a cellular microstructure composed of microgrooves that are not flat; rather they are convexly curved, enveloping an interconnected network of channels (inset, Fig. 1b). Here, we focus on the effect of such a convexly grooved microstructure on the plastron formation between the coarser fibrillar topography.

3D printed architectures with interconnected elliptical grooves are employed to study the formation of the SGL and its effect on plastron restoration. We characterize the response for varying surface curvatures, groove spacings and liquid pressures (Fig. 1b). Comparison with control experiments consisting of V-shaped grooves[12] allows us to directly quantify the curvature effect. Theoretical analyses and all-atoms computations of the SGL within individual grooves yield a fundamental understanding of its formation as well its interplay with the inter-groove spacing that sets the storage capacity of the plastron. Our results identify the material parameters and groove microstructures for the design of SHPOSs with universally stable and restorable gaseous plastrons.

**Results**

**Observations of SGL on individual grooves.** The individual grooves are slowly immersed in water to initiate the formation of an SGL. We apply a small hydrostatic pressure as the resultant pressure difference across the water meniscus allows us to visualize the liquid-gas (L-G) interface on the grooves. Fig. 1c-f shows the response of four pairs of elliptical and V-shaped grooves with varying depth $b$. The light bar (dashed line in Fig. 1c) represents the projection of the L-G interface on each groove and the dark bar is the reflection of light from the interface cross-section. The elliptical grooves lead to a stable SGL for all groove depths, while the corresponding V-shaped grooves stabilize the SGL at large groove depths with a consistently smaller layer thickness $h$. Specifically, for shallow V-shaped grooves with $b/a = 0.5$, the SGL is almost non-existent and the structure transitions to the Wenzel state (Fig. 1c).

**Plastron restoration experiments.** The elliptical grooves offer a distinct advantage in the stabilization of a thicker SGL, facilitating efficient gas transport through the interconnected grooves, thereby making full use of the replenishment to form a percolated gaseous film. Its beneficial effect should cascade over to the restoration of the plastron. To test this hypothesis, we perform experiments on dual-scale SHPOSs. Fig. 2a-b show the elliptical and V-shaped grooves ($b/a = 0.5$) at the base of one of the sixteen pillars within the as-fabricated SHPOSs. On slow immersion, the air column between the pillar arrays collapses under the hydrostatic pressure. For elliptical grooves, we observe a remnant SGL that is contiguous along the groove channel, visible as dark stripes in the top panel of Fig. 2c. In the case of the V-shaped grooves, the layer is unable to form completely. Rather, we see small irregular air pockets along the channel, visible as dark spots, consistent with the individual groove response (Fig. 1).

Fig. 2c-d show the subsequent air restoration response at a constant air injection rate of 0.02 mL/min (top panels) along with their schematic representations (bottom panels).

For elliptical grooves, the air rapidly replenishes the substrate through the interconnected SGL that evolves into a flat L-G interface. The air plastron grows uniformly in thickness along the pillars until it reaches the top of the pillar array to complete the restoration process. The entire process is schematically summarized in the bottom panel of Fig. 2c. The response of the V-shaped grooves is fundamentally different (Fig. 2d). The interaction between the discontinuous air pockets and the injected air results in formation of localized air bubbles that grow within the pillars without percolating throughout the surface. Notably, the bubble formation precludes the formation of a planar plastron. The top of the bubble gradually detaches from the pillar arrays and grows vertically until it escapes the SHPOS. As more air is injected, the bubbling is always favored and leads to unsuccessful air restoration, as illustrated in the bottom panel of Fig. 2d.

**Mechanisms of SGL on two-dimensional (2D) elliptical groove.** The experiments establish a clear link between stable SGLs and restorable plastrons. We quantify the fundamental basis for the stability of these layers using MD simulations of a water meniscus suspended between a quasi-2D elliptical groove within a compute cell of length 5.41 nm. Fig. 3a depicts the atomic configurations of the groove for two different initial positions of the meniscus at a constant (positive) Laplace pressure. In both cases, the liquid rapidly recedes or advances to a stable position, suggesting an equilibrium state likely set by the interplay between the groove shape, Laplace pressure and the liquid contact angle associated with the surface hydrophobicity. To identify the relevant parameters and their effect, we perform an energetic analysis based on the minimization of the total grand potential[28] for a liquid-gas-solid (L-G-S) three-phase system shown in Fig. 1b. The equilibrium conditions are (see SI for the derivation)

$$\Delta p r_0 - \gamma_{LG} = 0 \quad (1)$$

$$\theta_Y = \text{constant} \quad (2)$$

where $\gamma_{LG} = 0.0079 \text{ N/m}$[29] is the liquid surface tension. Eq. 1 is the Young-Laplace equation that is valid for both positive and negative Laplace pressures and Eq. 2 serves as a boundary condition[30] set by the constraint of the constant liquid contact angle. Combining the two, we get the condition for the radius of curvature of the meniscus $r_0 = r_0(\alpha_0)$

$$r_0 = \frac{a - \rho(\theta)\cos\theta}{\sin\varphi_0} \qquad (3)$$

with $\rho(\theta) = ab/\sqrt{b^2\cos^2\theta + a^2\sin^2\theta}$, $\theta = \theta(\alpha_0) = \arctan\left[\left(\frac{a^2}{b^2}\right)\tan\alpha_0\right]$, and $\varphi_0 = \alpha_0 + \theta_Y - \pi/2$. The tangent to the contact line $\alpha_0$ is the sole variable in Eq. 3 and represents the solution for the equilibrium position of the liquid-solid (L-S) contact line on the groove surface. The extrema $\alpha_0 = 0$ and $\alpha_0 = 90°$ correspond to the wetted (Wenzel) and fully gaseous states, respectively.

The solution $\alpha_0 \equiv \alpha_0(b/a, \Delta pa/\gamma_{LG}, \theta_Y)$ depends on the groove geometry $b/a$, the dimensionless Laplace pressure $\Delta pa/\gamma_{LG}$, and the liquid contact angle $\theta_Y$ set by the surface hydrophobicity. As validation, we perform MD simulations at a fixed groove spacing $a = 10$ nm. Fig. 3b-c shows the plots of equilibrium tangent angle $\alpha_0$ for the quasi-2D elliptical grooves as a function of $b/a$, $\Delta pa/\gamma_{LG}$, and $\theta_Y$, respectively. In each case, the simulation results are in excellent agreement with the theoretical predictions. For $\Delta p = 1$ atm and $\theta_Y = 135°$, there is negligible change in $\alpha_0$ for orders of magnitude variations in $b/a$ (top, Fig. 3b). We observe a slight decrease for shallow grooves with small $b/a$, approaching the limiting value $\alpha_0 = 37.9°$ as $b/a \to \infty$. Evidently, the L-S contact line is highly stable and allows the formation of the SGL for a wide range of groove geometries. The bottom panel in Fig. 3b shows the effect of positive Laplace pressures at fixed groove geometry $b/a = 1$. We observe a rapid decrease in $\alpha_0$ at low pressures and it stays positive to dimensionless pressures in excess of 6.5. The trend is consistent with the decrease in the meniscus curvature with increasing pressure (Eq. 1 and Eq. 3).

The effect of $\theta_Y$ is different for positive and negative pressures, as seen from the plot in Fig. 3c for $b/a = 1$. For positive pressures, hydrophilic surfaces ($\theta_Y < 90°$) are unable to sustain the SGL and the groove surface is completely wetted, as expected (left inset, Fig. 3c). For hydrophobic surfaces ($\theta_Y > 90°$), we see an almost linear increase in $\alpha_0(\theta_Y)$ to its maximum value at $\theta_Y = 180°$. It deviates slightly from the fully gaseous state, i.e. $\alpha_0(\theta_Y = 180°) < 90°$ as the positive pressure induces a finite meniscus curvature. Analogously, a negative pressure reverses the meniscus curvature that now aids the formation of the fully gaseous state for hydrophobic surfaces with $\theta_Y < 180°$. The slope of the curve $\alpha_0(\theta_Y)$ is larger than that for the linear increase at $\Delta p = 0$ (dashed line). The response of hydrophilic surfaces to negative pressures is qualitatively different. The transition to complete wetting is mediated by the formation and eventual escape of spherical gas bubbles with the wetting liquid trapped underneath (right inset, Fig. 3c). The solution indicated by the dotted line corresponds to the position of the liquid meniscus under the bubble and represents the classical Concus-Finn (C-F) condition for the wetting of an interior-corner, $\alpha_0 + \theta_Y < \pi/2$[12, 31]. Our solution deviates from the linear increase predicted by the C-F condition due to finite (negative) pressure. Above this line, there is no solution for the meniscus formation due to the geometrical constraint imposed by the groove convexity. The formation and escape of the unstable air bubbles is observed in air restoration experiments on the SHPOS without a hydrophobic coating (Fig. S3).

Fig. 3d shows the combined effect of the dimensionless pressure $\Delta p a/\gamma_{LG}$ and the hydrophobicity $\theta_Y$ as a contour plot of $\alpha_0$. The phase diagram naturally divides the $\theta_Y - \Delta p a/\gamma_{LG}$ space into four zones. For hydrophilic surfaces, the groove wets uniformly or by the nucleation and escape of gas bubbles. In both cases, the meniscus does not form and $\alpha_0 = 0$. For hydrophobic grooves, the groove convexity together with the pressure-dependent meniscus curvature stabilizes the SGL. The MD simulations and the theoretical solutions confirm that for convex, quasi-2D grooves, the hydrophobicity of the surface is the necessary condition for forming a stable SGL, which is

$$\theta_Y > 90° \qquad (4)$$

As a direct comparison, the black dashed line in Fig. 3d is the condition for V-shaped grooves, $\theta_Y > 90° + \arctan(a/b)$[12], indicating that the stable SGL formation for these classes of grooves is restricted to a narrower range of hydrophobic surfaces and deeper groove geometries.

**Three-dimensional (3D) groove.** The radial fluctuations of the liquid meniscus of the 3D grooves (Fig. 4a) can lead to additional instabilities[12, 31] and modify the phase diagram. To quantify their extent, we first perform MD simulations of the contact line within a 3D groove of axial length 72.56 nm. The simulations are performed at fixed pressure, groove geometry and hydrophobicity. The equilibrium tangent angle $\alpha_0$ is averaged within slices of the same length as the quasi-2D grooves (5.41 nm), as shown in Fig. 4b. Fig. 4c shows the plot of $\alpha_0$ along the groove length. The fluctuation amplitude is within a few degrees of value for the quasi-2D groove, also plotted for reference.

The wavy liquid meniscus is a consequence of the minimization of the liquid-gas interfacial energy and the constraints imposed by the groove geometry and the liquid pressure[31]. We then expect the extent of the fluctuations to scale with the length of the groove. Unstable fluctuations can lower the free energy of the three-phase system by triggering a collapse of the gas layer into smaller gas bubbles, thereby rendering the SGL unstable. We perform a stability analysis by introducing the mechanical free energy $F_{me} = \Delta p V_G + \gamma_{LG} A_0$[31-32], a simpler form that is equivalent to the total grand potential[32], where $V_G$ is the gas volume and $A_0 = A_{LG} + A_{SG}\cos\theta_Y$ is the effective area set by the L-G and S-G areas $A_{LG}$ and $A_{SG}$, respectively. For a radially fluctuating meniscus, the free energy is modified is due to changes in both the gas volume and the effective area $F'_{me} = \Delta p V'_G + \gamma_{LG} A'_0$. A net increase in the free energy $\delta F_{me} = F'_{me} - F_{me} > 0$ implies stable fluctuations of the meniscus about the 2D equilibrium state, while for $\delta F_{me} < 0$ the fluctuations grow and eventually destroy the seed gas column into smaller air pockets. In effect, the meniscus stability requires that the minimum free energy variation $(\delta F_{me})_{min} = (F'_{me})_{min} - F_{me}$ is positive.

We analyze the stability for radial perturbations of the meniscus of the form $r(\varphi, z) = \sum_k r_k(\varphi)\exp(ikqz)$, where $q \in (1, \infty)$ is the inverse wavelength of the perturbation (the wavenumber) and $\varphi$ and $z$ are the azimuthal and axial coordinates. The groove geometry together with the condition $(\delta F'_{me})_{min} > 0$ yields the stability relation for a meniscus fluctuating about $\varphi = \varphi_0$ (see SI for details)

$$\Delta p \left[s\tan(s\varphi_0) - \tan\left(\frac{\pi}{2} - \theta_Y\right)\right] > 0 \qquad (5)$$

where $s^2 = 1 - q^2 \in (-\infty, 1)$ captures the effect of the perturbations.

The $\theta_Y - \alpha_0$ and $\theta_Y - \Delta p a/\gamma_{LG}$ phase diagrams based on the additional stability criterion embodied by Eq. 5 are plotted in Fig. 4d-e, respectively. As before, the phase diagrams reveal four distinct regimes of behavior. The response of hydrophilic surfaces remains unchanged compared to the quasi-2D groove in that the liquid wets the groove for positive pressures while the C-F condition for unstable gas bubble formation still applies for negative pressures. For hydrophobic surfaces, the SGL is stable to the radial fluctuations for positive pressures and the solution for the meniscus is identical to that for the quasi-2D groove. Negative pressures, though, destabilize the SGL with respect to these fluctuations as the mechanical free energy is lowered, thereby favoring the collapse of the meniscus and the gas column. The formation of these bubbles is observed during early stages of the air injection experiments on the SHPOS that likely induce negative pressure within the SGL, yet the restoration is still successful (Fig. 2c).

To get a better understanding of this effect we perform longer time-scale MD simulations. The four inset figures in Fig. 4e show the atomic configurations of the sectional view through the center of the groove. For a fluctuating meniscus with $\Delta p > 0$, $\theta_Y > 90°$, we see a stable SGL marked by uniform dispersion of gas phase clusters along the groove length (upper right inset, Fig. 4e). In this region of the phase diagram, the response of the 3D groove is identical to the quasi-2D groove (Fig. 4c) and the meniscus is resistant to pressure fluctuations. The quasi-2D solutions therefore apply, and we use the results for subsequent analyses, specifically for the comparison of convex groove response with that of the V-shaped grooves.

Reversing the pressure ($\Delta p < 0$, $\theta_Y > 90°$) results in large fluctuations that ultimately lead to nucleation of gas pockets. These pockets are in direct contact with the groove surface and do not have a wetting liquid underneath, and they are interconnected by a thinner gaseous layer that is the remnant of a stable SGL (lower right inset, Fig. 4e). On continual air injection, an undulating yet fully formed gaseous layer forms that serves as precursor to successful air restoration. Note that the trapping of the SGL in the unstable zone is absent for V-shaped groove. In contrast, hydrophilic surfaces ($\Delta p < 0$, $\theta_Y < 90°$) facilitate the formation of disconnected air bubbles with a wetting liquid underneath that corresponds to the C-F condition for the gaseous escape (lower left inset, Fig. 4e). The SGL layer does not form, leading to unsuccessful restoration. Finally, for hydrophilic surfaces subject to positive pressures, the liquid completely wets the groove surface, corresponding to a classical Wenzel state (upper left inset, Fig. 4e).

**Elliptical groove versus V-shaped groove.** The gas storage capacity of the SHOPs is a measure of their superhydrophobicity and therefore equally important for the robustness of the plastron. To quantify the effect of the groove convexity on the SGL volume, we define the (normalized) capacity factor $wh/2ab$, where $w$ and $h$ are the width and thickness of the SGL. Fig. 5a and 5b show the effect of the groove geometry and the pressure on the gas capacity. The V-shaped groove response is also shown for comparison. For shallow grooves with $b/a \leq 1$ and at high (positive) pressures, the convex grooves again offer a clear advantage over the V-shaped grooves in their storage capacity. Increasing the groove depth beyond $b/a = 1$ leads to an abrupt increase in the capacity of V-shaped grooves and higher storage capacities. A similar trend is observed at low pressures. While the slower variations for convex grooves limit their storage capacity, they confer superior pressure fluctuation resistance. The storage capacities extracted in MD simulations for quasi-2D grooves are also shown in the plots. Both the effect of groove geometry and pressure are in excellent agreement with the theoretical predictions.

Fig. 5c-d combines these trends within a phase diagram of the SGL volume $wh/2ab$ for the elliptical and V-shaped grooves at fixed groove geometry $b/a = 1$, respectively. In both plots, the dashed line separates the wetted Wenzel state from the Cassie-Baxter state with a stable SGL. The superior storage capacity of convex grooves to large pressures is reinforced by the (geometric) stabilization of the SGL to a much larger range of hydrophobicity $\theta_Y \geq 90°$, considerably enhancing the robustness of the plastron formation and restoration.

**Improvements of SGL storage capacity of elliptical groove.** Hierarchical groove microstructures can also be engineered with independent control over the groove spacing, thereby amplifying their gas storage capacity. Fig. 6a is a schematic illustration of the design employed for studying the SGL formation within such architectures. The separated groove supports a suspended liquid and its limit is set by groove spacing at which the liquid sags and makes mechanical contact with the groove bottom. This triggers the onset of the Cassie-Baxter-Wenzel transition as the energetic barrier for wetting of the surface underneath the liquid is negligible[33], resulting in the loss of the SGL.

The dimensionless maximum groove separation $x/a$ corresponding to the critical point at which the meniscus contacts the groove bottom can be expressed as a condition that the height of the SGL layer is zero

$$\rho(\theta)\sin\theta - r_0(1 - \cos\varphi_0) = 0 \qquad (6)$$

where $r_0 = (a + x - \rho(\theta)\cos\theta)/\sin\varphi_0$. The first term in Eq. 6 is the height of the contact line and the second term is the meniscus depth. Using Eq. 1 and 2, we numerically solve

Eq. 6 to extract the critical separation $x/a$ under varying conditions. The results are plotted in Fig. 6b-d as a function of $b/a$, $\Delta pa/\gamma_{LG}$, and $\theta_Y$, respectively. For each plot, we also perform MD simulations of the meniscus suspended between quasi-2D grooves with varying inter-groove distances. Fig. 6b shows the effect of the groove geometry on the critical separation $x/a$ at fixed pressure and hydrophobicity ($\Delta p = 1$ atm, $\theta_Y = 135°$). $x/a$ increases non-linearly and the trend is in good agreement with the observations within MD simulations. As $b/a$ approaches infinity, we recover the solution for a high aspect ratio rectangular groove, *i.e.* $x/a = r_0 \cos\gamma$, or $x/a = 5.51$ for the chosen pressure and hydrophobicity. This extreme case corresponds to a fibrillar groove microstructure, and the SGL is stable as the liquid meniscus assumes a height larger than the meniscus depth.

The pressure dependence shown in Fig. 6c is an inverse variation in $x/a$ at fixed groove geometry and hydrophobicity ($b/a = 1$, $\theta_Y = 135°$). As the pressure approaches zero, the groove separation can be arbitrarily large in the absence of gravity as the meniscus is devoid of any curvature. Increasing pressure changes both the meniscus curvature and the meniscus height, leading to a drastic initial decay in $x/a$. Thereafter, the decay is slower as change in meniscus height tapers off due to the groove convexity. The rapid initial decay indicates that in this regime, both the stability of the SGL and its storage capacity in separated grooves is increasingly susceptible to pressure fluctuations. The effect of hydrophobicity plotted in Fig. 6c shows that increasing $\theta_Y$ always enhances the groove separation, as expected, and it asymptotes to a constant for the limit $\theta_Y = 180°$. MD simulations of both the pressure and hydrophobicity variations are in agreement with the theoretical predictions.

**Discussions**

Our studies on individual grooves show that the finite (convex) groove curvature forces the interfacial balance that sets the location of the liquid meniscus to adapt to the groove geometry. The geometric degree of freedom enables the groove microstructure to trap and stabilize an SGL for wide variations in the (positive) Laplace pressures typically encountered in underwater applications. Crucially, the SGL is stable for *all* hydrophobic surfaces, eliminating the need for aggressive surface treatments that are often needed to ensure a high degree of hydrophobicity. Plastron formation and restoration experiments within dual-scale structures conclusively show that this added stability of the SGL layer leads to robust gaseous plastrons that are otherwise susceptible to pressure fluctuations. For interconnected groove microstructures, local collapse of the plastron is effectively countered by restoration through the SGL, eliminating the need for external stimuli.

Integration of this design principle within hierarchical architectures enables the surfaces to preserve their superhydrophobic state underwater and is likely the basis for the leaf microstructure observed in *Salvinia*. The microstructural parameters associated with the surface grooves on the leaves are $a \cong 13\ \mu$m and $b/a \cong 1$ (inset, Fig. 1b). For typical aquatic environments with $\Delta p = 1 - 2$ atm, the dimensionless pressures are in the range $\Delta p a/\gamma_{LG} = 20 - 40$. Under these conditions, the SGL is always stable underwater with a higher gas volume relative to V-shaped grooves (Fig. 5), facilitating the formation of a stable gaseous plastron for a wide range of aquatic environments. For nanoscopic hierarchical structures with sub-micron scale groove spacing such that $\Delta p a/\gamma_{LG} \approx 1$, the effect is more dramatic; the SGL is stable with a higher gas capacity that results in improved pressure fluctuation resistance.

In general, the higher SGL gas capacity represents a measure of the robustness of hydrophobicity of the surface. The V-shaped geometries are superior, provided the grooves are consistently deep enough as they are unable to support an SGL for grooves with convexities $b/a < 1$ (Fig. 5a). The convexity enables shallow grooves to be equally effective while improving the pressure fluctuation resistance for a wider range of liquid-surface systems, simplifying their design and scalable synthesis. Additionally, the groove convexity in tandem with a separated groove design enhances their gas storage capacity, suggesting that optimization of the interplay between the groove convexity and separation can be used to tune their overall robustness.

For hydrophilic surfaces, V-shaped grooves can stabilize discrete bubbles and prevent complete wetting for geometries and surfaces that satisfy the relation $\arctan(a/b) + \theta_Y > \pi/2$[12]. Convex grooves are unable to prevent wetting as hydrophilic surfaces trap the liquid beneath the discrete bubbles that eventually escape. However, the convex geometry can be combined with cavity-based designer surface roughness[34-35]. Optimization of the surface morphologies at different length-scales opens up the possibility of robust omniphobic surfaces in a broad range of material systems within emerging digital manufacturing frameworks.

**Materials and Methods**

**Sample Fabrication.** 3D computer-aided design software (Solidworks) is used to design the individual groove samples as well as the dual-scale SHPOSs composed of an array of cylindrical pillars supported on interconnected grooves. The elliptical grooves (width × length × height = 2400 μm × 4800 μm × 3000 μm) are fabricated with varying surface curvatures, parametrized by their axes (depth-width) ratio $b/a$. For each groove geometry, a corresponding control sample with V-shaped grooves with the same depth $b$ is also fabricated. The cylindrical pillars within the 9.6 mm × 9.6 mm SHPOS samples are of diameter 800 μm and height 2000 μm and placed above the grooves arranged into a 4 × 4 array with an inter-pillar distance of 2400 μm.

All samples are fabricated with the negative photoresist using a 3D projection micro-lithograph stereo exposure system (S130, BMF Material Technology Inc. Guangdong province, China). The sample surfaces are treated with a commercial hydrophobic solution (Glaco, Japan) to achieve an intrinsic water contact angle of $141 \pm 2.3°$ (ImageJ software, Fig. S4).

**Immersion experiments.** (1) *Individual grooves*: The elliptical and V-shaped grooves are glued to a sliver aluminum plate with a waterproof double-sided tape and submerged underwater in a transparent sealed plexiglass box. The topside of the box serves as the observing window. (2) *SHPOSs*: The samples are glued to a black aluminum plate and submerged underwater in the same plexiglass box. The box sidewall is used for observation. Scanning electron microscopy (Genimi SEM 500, Carl Zeiss Microscopy Ltd. Germany) is used to characterize the interfaces within the SHPOS structures at a voltage of 3 kV. A syringe pump (LSP02-2A, Longer, China) with its needle tip placed adjacent to the SGL replenishes the air between the pillars. A digital microscope (AM4113T, Dino-Lite, China) is used to observe the SGL and the air restoration process in both experiments.

**Computations.** All-atom molecular dynamics (MD) simulations of the suspended liquid meniscus within individual grooves are performed using LAMMPS. The isothermal, isostress (NP$_{ZZ}$T) ensemble allows us to vary the liquid pressure $p_L$ along the surface normal. The gas phase is assumed to be the vapor phase[36-37] at the liquid saturation vapor pressure[38]. The stability of the gas phase is studied for both positive and negative Laplace pressure ($\Delta p = p_L - p_G$). The latter can arise during restoration and is modeled by increasing the gas mass within the trapped SGL. Gas dissolution is small and takes place over times scales much longer than that for mechanical equilibration, and therefore neglected.

The solid is modeled as an FCC single crystal with a lattice constant of 5.13 Å. The Nosé-Hoover thermostat with a time step of 5 fs is used to control the temperature at $T = 300$ K. Periodic boundary conditions are employed in all three directions. A standard Leonard-Jones (LJ) potential (bond strength $\varepsilon$ and characteristic length scale $\sigma$) with cutoff radius $2.5\sigma$ is employed to model all pairwise atomic interactions within the solid (S), liquid (L) and gas (G) phases (Table S1).

The bond strength between the liquid and solid $\varepsilon_{LS}$ is tuned for varying intrinsic liquid contact angles on the solid, extracted from multiple equilibrium configurations using ImageJ. The value of $\varepsilon_{LS}$ and their corresponding contact angles are tabulated in Table S2. A least-square fit to our data plotted in Fig. S5 shows that $\cos\theta_Y$ increases linearly with $\varepsilon_{LS}$

$$\cos\theta_Y = 3.88\varepsilon_{LS} - 1.41$$

consistent with past LJ computations[39-40].

**Acknowledgments**

**Funding:** This work was supported from the National Natural Science Foundation of China (Grant No. 12172347) and the Fundamental Research Funds for the Central Universities (Grant No. WK2480000006). The authors are grateful for supercomputing resources available through the Massachusetts Green High Performance Computing Center (MGHPCC).

**Author contributions:** X.H., M.U., and H.W. conceived and designed the study. X.H. performed all the experiments and simulations. X.H., J.L., M.U., and H.W. analyzed the data and discussed the results. H.W. and M.U. supervised the work. X.H., M.U., and H.W. wrote the paper.

**Competing interests:** Authors declare that they have no competing interests.

**Data and materials availability:** All data needed to evaluate the conclusions in the paper are present in the paper and/or the Supplementary Materials. Additional data related to this paper may be requested from the authors.

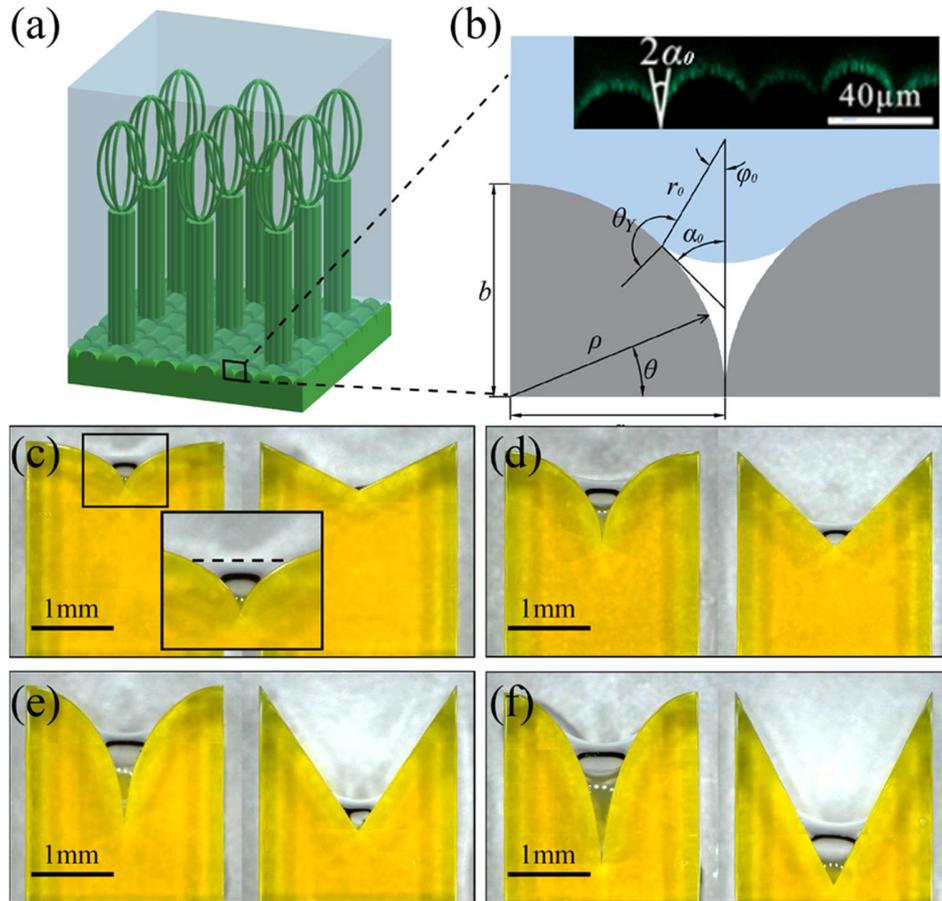

**Figure 1. Experiments on SGL formation within individual grooves.** (a) 3D sketch of an immersed *Salvinia* leaf illustrating the dual-scale surface structure composed of the eggbeater-shaped fibrillar outgrowths supported by a grooved substrate. (b) Schematic representation of the immersed grooves of depth $b$ and half-width $a$ with a suspended water meniscus and (inset, b) the confocal image of curved profile of the cross-section of microgrooves. Reproduced with permission[12]. Copyright 2020, National Academy of Sciences. Here, $\alpha_0$ is the L-S contact line tangent angle at the groove, $\theta_Y$ is the liquid contact angle, $r_0$ is the radius of curvature of the meniscus, $\varphi_0$ is its azimuthal coordinate, and $\rho$ and $\theta$ are the radial and angular coordinates of the groove surface, respectively. (c-f) Observed projection of the L-G interfaces (light lines) for elliptical (left) and V-shaped (right) grooves with depth-width ratio (c) $b/a = 0.5$, (d) $b/a = 1$, (e) $b/a = 1.5$, and (f) $b/a = 2$. (inset, c) Magnified view of the elliptical groove with the L-G interface marked by the dashed line.

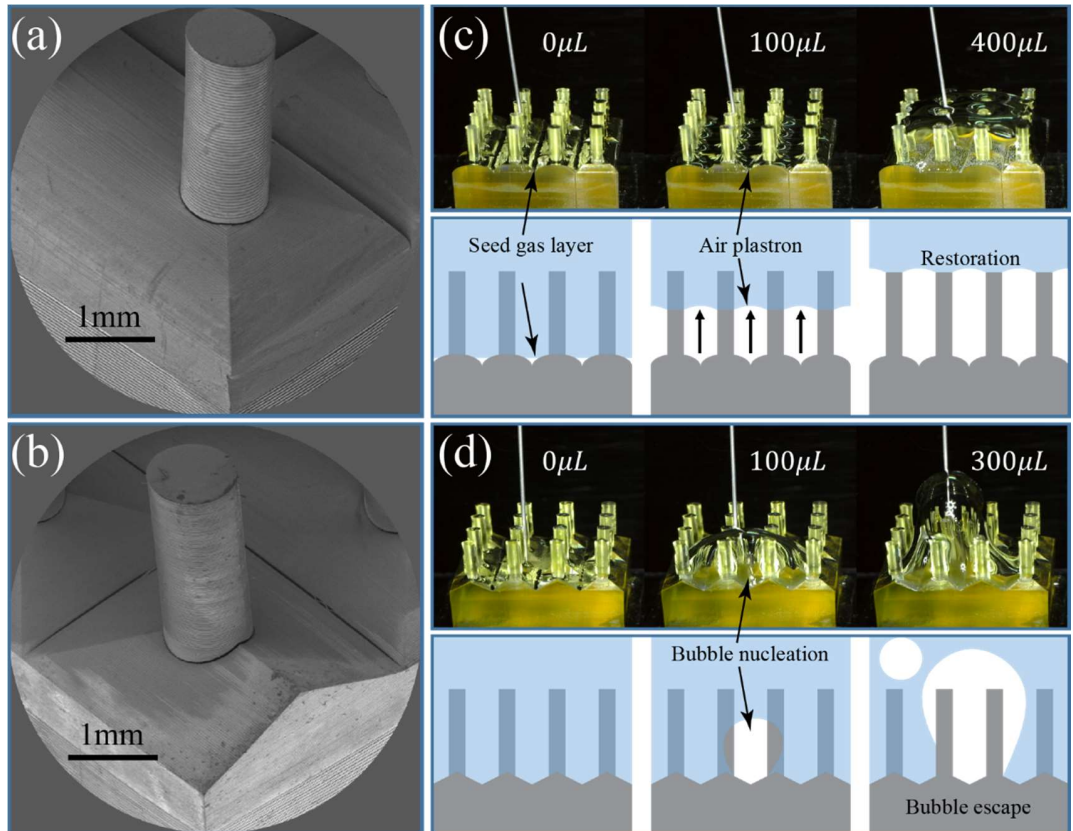

**Figure 2. Plastron restoration experiments on dual-scale SHPOSs.** (a-b) SEM micrographs showing the (a) elliptical and (b) V-shaped surface grooves in the vicinity of a cylindrical pillar in dual-scale SHPOS samples. The depth-width ratio $b/a = 0.5$ for both classes of grooves. (c-d) Image sequence showing (top panel, c) successful air restoration in the SHPOS covered by elliptical grooves with (bottom panel, c) its schematic illustration and (top panel, d) unsuccessful air restoration for the SHPOS covered by V-shaped grooves with (bottom panel, d) its schematic illustration. The air injection rate is held constant at 0.02 mL/min.

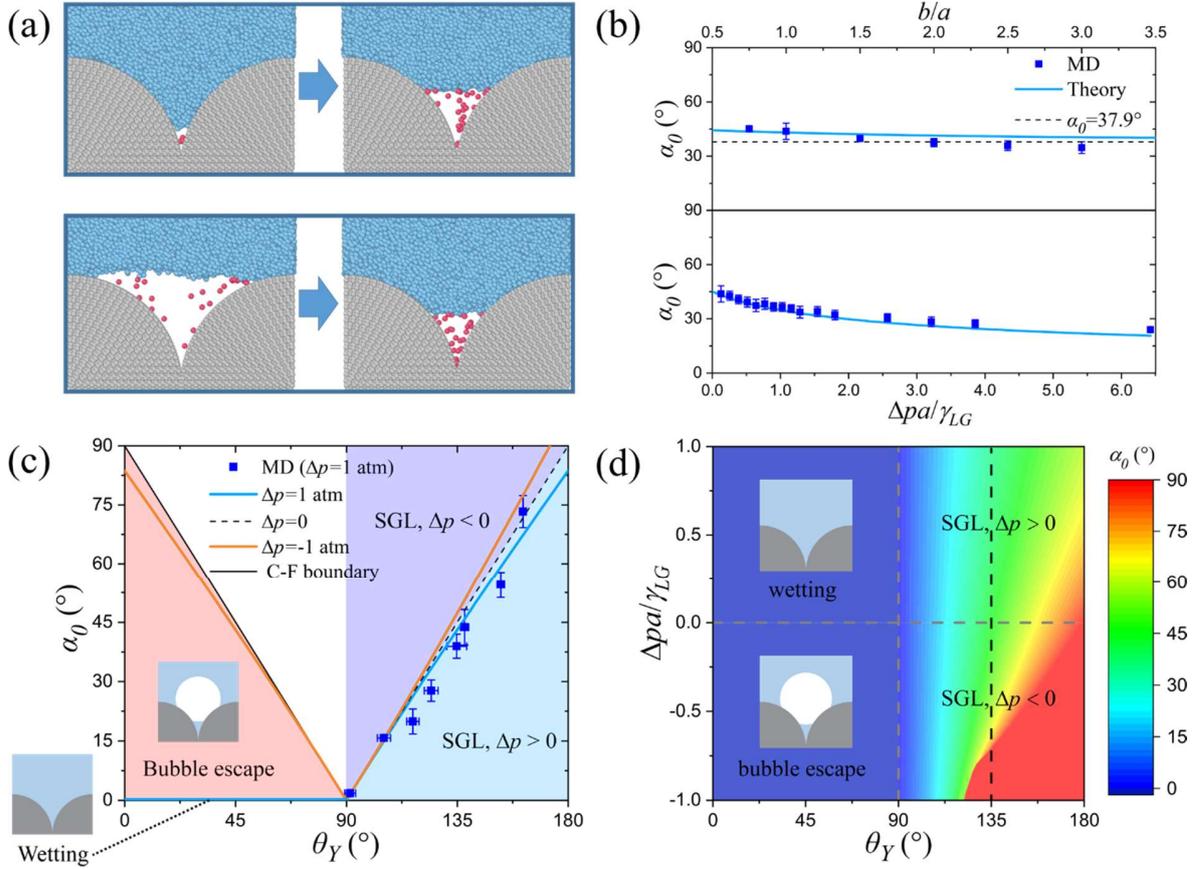

**Figure 3. SGL formation within quasi-2D elliptical grooves.** (a) Atomic configurations within MD simulations of the spontaneous recession (top panel) and advancement (bottom panel) of the meniscus on a cylindrical groove for two different initial positions of the contact line. Light blue, light red, and grey indicate liquid, gas, and solid, respectively. For consistency, the color scheme for the atomic plots is preserved in the remaining figures. (b) Contact line position parametrized by tangent angle $\alpha_0$ (Fig. 1b) for various groove geometries $b/a$ (top panel) and dimensionless pressures $\Delta p a/\gamma_{LG}$ (bottom panel), respectively. The lines and solid dots represent the theoretical solutions and MD results, respectively. Here and elsewhere, the simulations and theoretical solutions correspond to reference values of width $a = 10$ nm, $b/a = 1$, $\Delta p = 1$ atm, and $\theta_Y = 135°$. (c) Tangent angle $\alpha_0$ for various liquid contact angles $\theta_Y$ with Laplace pressure $\Delta p = \pm 1$ atm, respectively. (d) Phase diagram shown as a contour plot of $\alpha_0$ as a function of dimensionless pressure $\Delta p a/\gamma_{LG}$ and liquid contact angle $\theta_Y$.

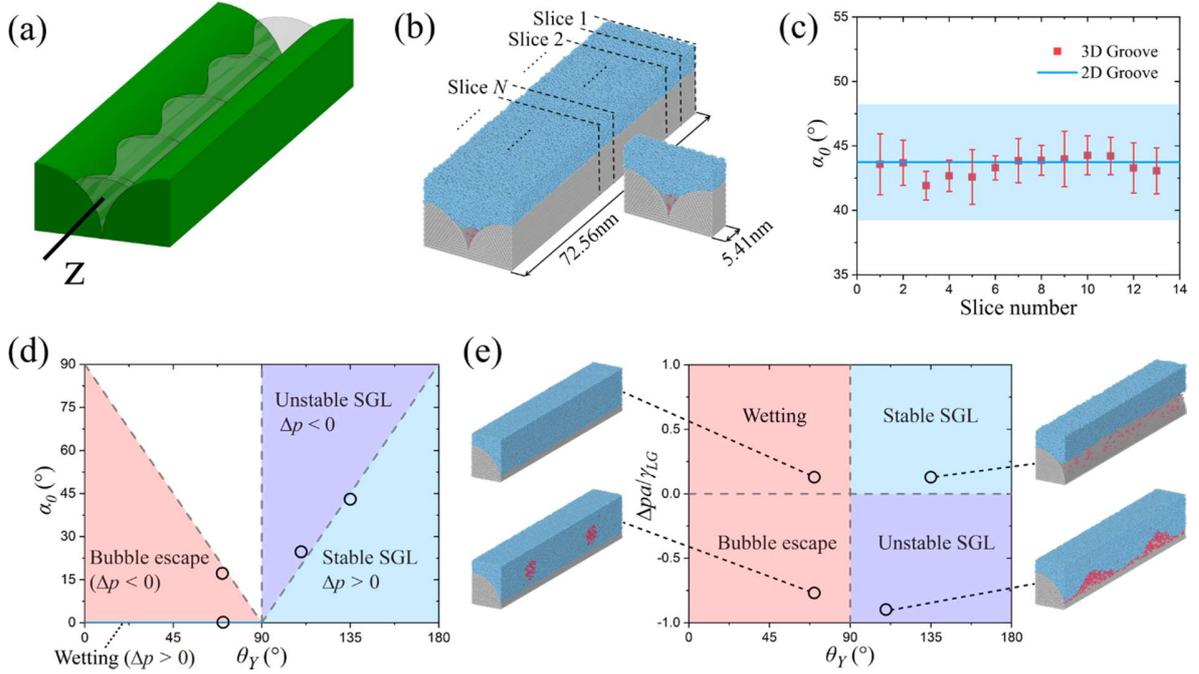

**Figure 4. SGL stability within 3D groove columns.** (a) Schematic representation of the fluctuating liquid meniscus on a cylindrical groove, where the transparent part represents the gas. (b) MD configurations of the 3D (left) and 2D (right) cylindrical grooves. Here and elsewhere, the plots are for reference values of the groove spacing $a = 10$ nm, liquid contact angle $\theta_Y = 135°$, and Laplace pressure $\Delta p = 1$ atm if not specified. (c) The equilibrium tangent angle $\alpha_0$ extracted using MD simulations along the 3D groove length (dots). The value for the quasi-2D groove is also shown (line). (d-e) Phase diagram of the stability of the liquid meniscus to radial fluctuations in the (d) $\theta_Y - \alpha_0$ space and (e) $\theta_Y - \Delta p a/\gamma_{LG}$ space, respectively, with sectional view of the MD snapshots of the cases in the (upper right inset, e) stable regime, (lower right inset, e) unstable regime, (lower left inset, e) bubble formation and escape regime, and (upper left inset, e) wetting regime.

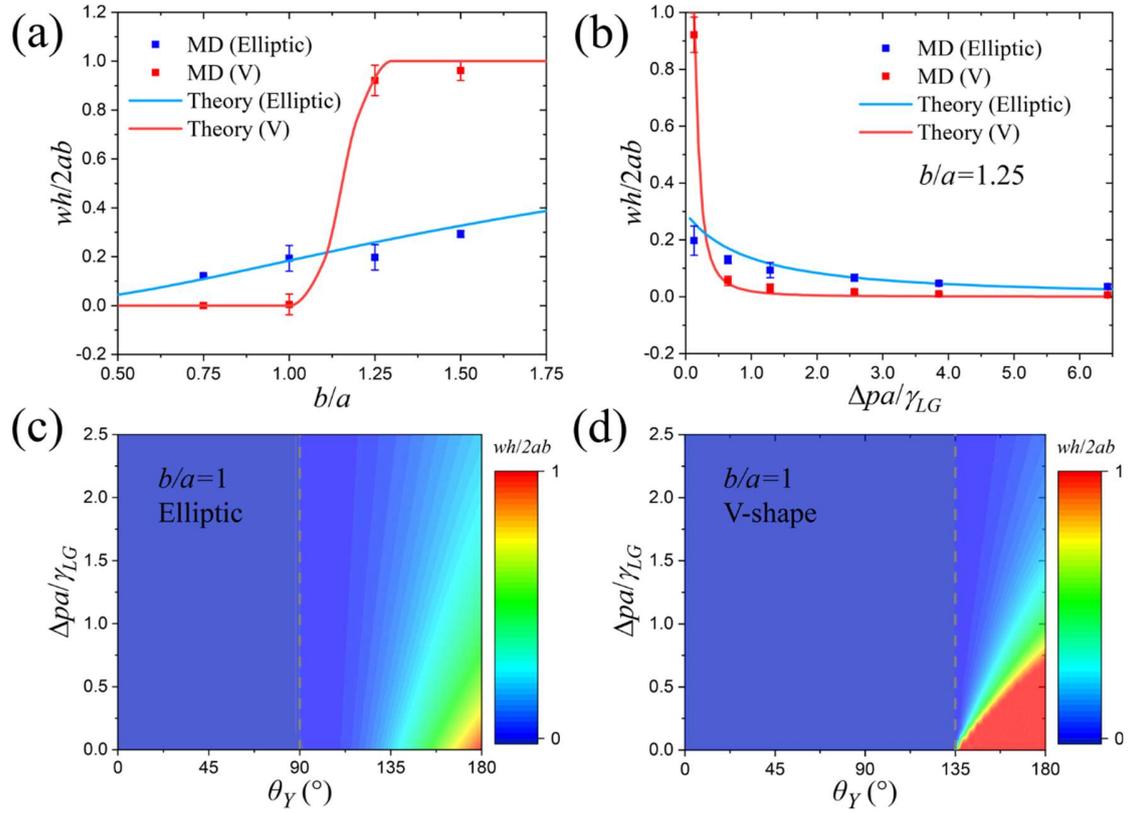

**Figure 5. Effect of convexity on groove gas storage capacity.** (a-b) Gas storage capacity of the two classes of grooves for different (a) geometries $b/a$ and (b) dimensionless pressures $\Delta p a/\gamma_{LG}$, respectively. (c-d) Contour plot of gas storage capacity factor $wh/2ab$ as a function of $\Delta p a/\gamma_{LG}$ and $\theta_Y$ for the (c) elliptical and (d) V-shaped grooves, respectively. The dashed line represents the condition for stable SGL formation. Solid lines represent the theoretical solutions and dots represent the MD results.

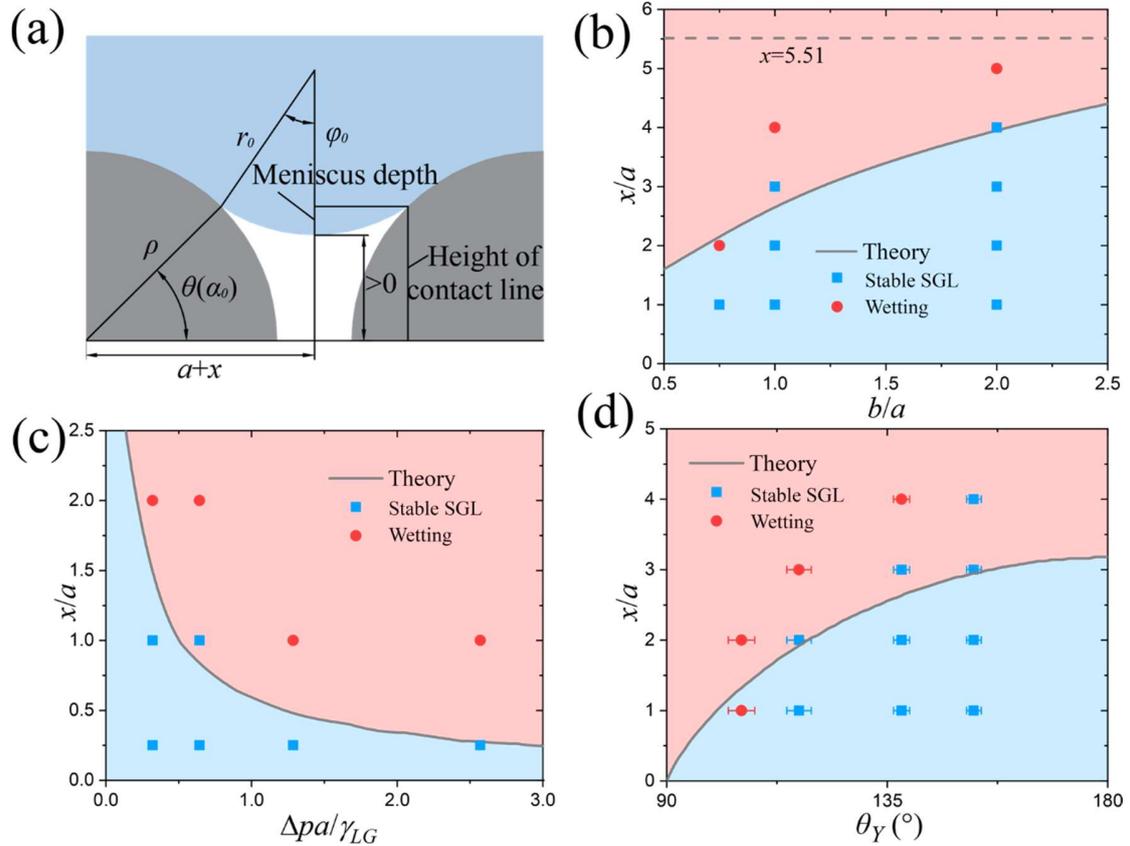

**Figure 6. SGL stability within separated elliptical grooves.** (a) Schematic illustration of a liquid meniscus suspended between a separated elliptical groove. (b-d) Stability phase diagrams plotted as a function of the dimensionless maximum separating distance $x/a$ and the (b) groove geometries $b/a$, (c) dimensionless pressures $\Delta p a/\gamma_{LG}$, and (d) liquid contact angles $\theta_Y$, respectively. The light blue zones indicate a stable liquid meniscus suspended between the grooves, while the light red zones correspond to wetting via mechanical contact with the groove bottom. The blue and red dots are the results of the MD simulations.

**Supplementary Material**

**1. 2D convex grooves**

We derive **Eqs. 1** and **Eq. 2** in the main text using the total grand potential and the Gibbs excess contribution for a liquid-gas-solid (L-G-S) three-phase system (*S1*):

$$F_{tot} = -p_L V_L - p_G V_G + \gamma_{LG} A_{LG} + \gamma_{SG} A_{SG} + \gamma_{SL} A_{SL},$$

where $p_L$, $p_G$, $V_L$, and $V_G$ represent the liquid pressure, gas pressure, and their corresponding volumes, respectively. $\gamma_{LG}, \gamma_{SG}, \gamma_{SL}$ and $A_{LG}, A_{SG}, A_{SL}$ denote the surface tension of the corresponding interfaces and their surface areas, respectively. Minimization of the functional $F_{tot}$ (i.e. $\delta F_{tot} = 0$) yields the equilibrium (stable) state of the system, and it depends on the variations in $\delta A_{LG}, \delta A_{SG}, \delta A_{SL}, \delta V_L$, and $\delta V_G$. Below, we analyze them individually.

Following **Fig. S1**, for the 2D elliptical groove, the liquid meniscus area $A_{LG} = 2\varphi_0 r_0$ and its variation $\delta A_{LG}$ is

$$\delta A_{LG} = 2(-\varphi_0 \delta r_0 + r_0 \delta \varphi_0).$$

The " $-$ " sign implies that radial and angular variations have different signs. The total solid surface area and the system volume are constant, i.e., $\delta A_{SG} = -\delta A_{SL}, \delta V_G = -\delta V_L$. The variation in the solid-gas area $\delta A_{SG}$ is simply

$$\delta A_{SG} = 2\rho \delta\theta.$$

The gas volume $V_G$ can be obtained by integrating $A_{LG}$ over the radial displacement $r$, which is $V_G = \int_0^{r_0} 2\varphi' r' \, dr'$. Hence, its variation at position $r_0$ is

$$\delta V_G = 2\varphi_0 r_0 \delta r_0.$$

Combining these relations, the variation of the functional $F_{tot}$ becomes

$$\delta F_{tot} = 2(p_L - p_G)\varphi_0 r_0 \delta r_0 + 2\gamma_{LG}(-\varphi_0 \delta r_0 + r_0 \delta \varphi_0) + 2(\gamma_{SG} - \gamma_{SL})\rho \delta\theta.$$

In the limit of infinitesimally small variations, $\delta\theta$ and $\delta\varphi_0$ are geometrically related, that is

$$\rho \delta\theta \cos\gamma = r_0 \delta\varphi_0,$$

where $\gamma = \pi - \theta_Y$ is the gas contact angle. Rearranging, we get

$$\delta F_{tot} = 2[(p_L - p_G)r_0 - \gamma_{LG}]\varphi_0 \delta r_0 + 2(-\gamma_{LG}\cos\theta_Y + \gamma_{SG} - \gamma_{SL})\rho\delta\theta.$$

Stable equilibrium $\delta F = 0$ then leads to the following relations:

$$(p_L - p_G)r_0 - \gamma_{LG} = 0, \quad\quad (S1)$$

$$\gamma_{SG} - \gamma_{SL} - \gamma_{LG}\cos\theta_Y = 0. \qquad (S2)$$

The two equations are in effect the Young-Laplace equation that relates the constant liquid contact angle to the interfacial energies, together with the boundary condition related to the convex groove geometry.

## 2. 3D grooves

### 2.1 Equilibrium geometry of 3D liquid meniscus

For simplicity, we ignore gravity and analyze cylindrical groove columns as shown in Fig. 4. Following the Gaussian method (S2-S4), we express the total free energy $E$ of the L-G-S system (S4) in terms of the total meniscus surface energy with the constant volume constraint:

$$E = \gamma_{LG} \iint \sqrt{1 + \left(\frac{r_{\varphi'}}{r}\right)^2 + (r_z)^2}\, r\, \mathrm{d}\varphi' \mathrm{d}z - \lambda \iint \frac{1}{2} r \cdot r\, \mathrm{d}\varphi' \mathrm{d}z,$$

where $r$, $\varphi'$, and $z$ are the radial, azimuthal, and axial coordinate of the liquid meniscus respectively, and $\lambda$ is a Lagrangian multiplier that enforces the constant volume constraint. Additionally, $r$ is a function of $\varphi'$ and $z$, and $\varphi'$ is a function of $z$, i.e., $r = r(\varphi', z)$ and $\varphi' = \varphi'(z)$. $\varphi'(z)$ reaches its maximum $\varphi = \varphi(z)$ at the L-S contact line. Then, the axial coordinate $z$ is the only variable in our theory. The equilibrium meniscus shape $\delta E = 0$ follows from the Euler-Lagrangian equation:

$$\frac{\mathrm{d}L}{\mathrm{d}r} - \frac{\mathrm{d}}{\mathrm{d}\varphi'}\frac{\partial L}{\partial r_{\varphi'}} - \frac{\mathrm{d}}{\mathrm{d}z}\frac{\partial L}{\partial r_z} = 0,$$

where $L$ is the Lagrangian function:

$$L = \gamma_{LG} r \sqrt{1 + \left(\frac{r_{\varphi'}}{r}\right)^2 + (r_z)^2} - \frac{\lambda}{2} r^2.$$

We take the Laplace pressure $\Delta p = p_L - p_G$ (pressure difference across the liquid meniscus) as the Lagrangian multiplier. Then, the Euler-Lagrangian equation becomes

$$\frac{\Delta p}{\gamma_{LG}} = \frac{1}{r\sqrt{1 + \left(\frac{r_{\varphi'}}{r}\right)^2 + (r_z)^2}}$$

$$- \frac{\left(\frac{r_{\varphi'}}{r}\right)_{\varphi'}(1 + (r_z)^2) - \frac{1}{r}r_{\varphi'}r_z r_{z\varphi'} - r_{\varphi'}r_z\left(\frac{r_{\varphi'}}{r}\right)_z + r r_{zz}\left(1 + \left(\frac{r_{\varphi'}}{r}\right)^2\right)}{r(1 + \left(\frac{r_{\varphi'}}{r}\right)^2 + (r_z)^2)^{\frac{3}{2}}}.$$

To make analytical progress, we restrict ourselves to bilinear terms by performing a Taylor expansion to linear terms in $r_\varphi/r$ and $r_z$. The equation simplifies to

$$1 = \frac{1}{r}\left(1 + \frac{1}{2}\left(\frac{r_{\varphi\prime}}{r}\right)^2 - \frac{1}{2}(r_z)^2\right) - \frac{1}{r^2}r_{\varphi'\varphi'} - r_{zz}, \qquad (S3)$$

Here, $r$ and $z$ are nondimensionalized by dividing them by $r_0$, the cylindrical meniscus curvature at the initial equilibrium state with no fluctuation.

## 2.2 Fluctuations of the 3D liquid meniscus

Following classic solution in **Ref. S4**, we expand $r$ as a Fourier series with axial wavenumber $q$:

$$r(\varphi', z) = \sum_k r_k(\varphi')\exp(ikqz).$$

Substituting into **Eq. S3** and retaining terms to second-order, we get

$$r(\varphi', z) = 1 + 2d_1\cos(s\varphi')\cos(qz) - \frac{1}{2}d_1^2[(1-2s^2)\cos(2s\varphi') + 1]\cos(2qz)$$

$$+ \frac{1}{2}d_1^2[\cos(2s\varphi') + (1-2s^2)], \qquad (S4)$$

where $d_1 \ll 1$, $s^2 = 1 - q^2 \in (-\infty, 1)$ for $q \in (1, \infty)$, and the dimensionless radius of curvature $r_0$ is 1 for convex meniscus shown in **Fig. S2a**. Comparison with the solution for the V-shaped groove in **Ref. S4** shows that the expression for $r(\varphi', z)$ does not depend on the shape of the groove if not written in its explicit form $r(z)$. That is, the groove shape only matters for the expression $\varphi = \varphi(z)$, which is also the boundary condition for the L-S contact line.

To solve for the $\varphi$ at the L-S contact line on the cylindrical groove, we use an additional geometrical relation for the length-scale $h$ shown in **Figs. S2a** and **S2b**:

$$r\sin(\varphi - \alpha) = \left[\frac{h_0}{\sin\alpha_0} - y_0(\cos(\alpha - \alpha_0) - 1) - x_0\sin(\alpha - \alpha_0)\right.$$

$$\left. - (x_0(1 - \cos(\alpha - \alpha_0)) + y_0\sin(\alpha - \alpha_0))\tan\alpha\right]\sin\alpha = h,$$

where

$$\alpha - \alpha_0 = \frac{(r - r_0)}{\rho}\cos\gamma,$$

and

$$h_0 = \sin\left(\gamma - \frac{\pi}{2}\right).$$

Using the geometrical relation $\rho\delta\theta\cos\gamma = r_0\delta\varphi_0$ (**Fig. S1**), the relation between the angles $\gamma - \pi/2 = \varphi_0 - \alpha_0$ (**Fig. S2a**) and **Eq. S4**, we get an explicit equation for $\varphi - \varphi_0$:

$$\sin\left(\gamma - \frac{\pi}{2}\right) - 2d_1\cos(s\varphi_0)\cos(qz)\left[\sin\left(\gamma - \frac{\pi}{2}\right)\right.$$
$$\left. - \cos\gamma\left(\frac{h_0\cos\alpha_0}{\rho\sin\alpha_0} - \frac{x_0\sin\alpha_0}{\rho} - \frac{y_0\tan\alpha_0\sin\alpha_0}{\rho}\right)\right]$$
$$= \sin\left(\gamma - \frac{\pi}{2}\right) + \left(\gamma - \frac{\pi}{2}\right)\cos\left(\gamma - \frac{\pi}{2}\right)$$
$$- \frac{2d_1\cos(s\varphi_0)\cos(qz)}{\rho}\cos\left(\gamma - \frac{\pi}{2}\right)\cos\gamma,$$

where we have used a series of Taylor series expansion of the trigonometric terms to first order. Rearranging, we arrive at an expression for the geometry of the fluctuating meniscus, parameterized by the following expression for $\varphi$:

$$\varphi = \varphi_0 - 2d_1\cos(s\varphi_0)\cos(qz)\left[\tan\left(\gamma - \frac{\pi}{2}\right) + C\right],$$

with the term

$$C = -\frac{h_0\cos\alpha_0\cos\gamma}{\rho\sin\alpha_0\cos\left(\gamma - \frac{\pi}{2}\right)} + \frac{x_0\sin\alpha_0\cos\gamma}{\rho\cos\left(\gamma - \frac{\pi}{2}\right)} + \frac{y_0\tan\alpha_0\sin\alpha_0\cos\gamma}{\rho\cos\left(\gamma - \frac{\pi}{2}\right)} - \frac{\cos\gamma}{\rho}$$

set entirely by the groove geometry.

## 2.3 Variational analysis of the mechanical free energy

The meniscus geometry forms the basis for the change in mechanical free energy of the system (*S4-S5*):

$$\delta F_{me} = \Delta p\delta V_G + \gamma_{LG}(\delta A_{LG} - \delta A_{SG}\cos\gamma),$$

where the changes in the gas volume, L-G interface area, and S-G interface area are calculated with respect to the meniscus with no fluctuations. The volume change can be written as

$$\delta V_G = \int\int_0^{\varphi(z)} r^2 d\varphi' + A_r dz - V_G,$$

where the gas initial volume

$$V_G = L\Pi = L\left[r_0{}^2\varphi_0 + \left(2\left(y_0 + \frac{h_0}{\sin\alpha_0}\right) + r_0\cos\varphi_0\right)r_0\sin\varphi_0 + \rho^2\sin\alpha_0\cos\alpha_0 - \rho^2\alpha_0\right].$$

The integral $\int_0^{\varphi(z)} r^2 d\varphi'$ is the sector enclosed by the liquid meniscus and the tangents to the L-S contact lines on the groove, and

$$A_r = \left(2\left(y_0 + \frac{h_0}{\sin\alpha_0}\right) + r\cos\varphi\right)r\sin\varphi + \rho^2\sin\alpha\cos\alpha - \rho^2\alpha$$

is the remainder of the cross-sectional area of the meniscus and the cylindrical groove. Integrating the cross-sectional area over a period length $L = 2\pi/q$ along the axial direction, the change in the gas volume is

$$\delta V_G = Lr_0{}^2\left[2d_1{}^2(1-s^2)\left(\varphi_0 + \frac{\sin(2s\varphi_0)}{2s}\right)\right.$$
$$+ 2d_1{}^2\left(s\tan(s\varphi_0) - \tan\left(\gamma - \frac{\pi}{2}\right)\right)\cos^2(s\varphi_0)$$
$$\left. - 4d_1{}^2\left(\tan\left(\gamma - \frac{\pi}{2}\right) + C\right)\cos^2(s\varphi_0)\right].$$

The change in the effective surface area $\delta A_0 = \delta A_{LG} - \delta A_{SG}\cos\gamma$ is

$$\delta A_0 = 2\int\int_0^{\varphi(z)} r_0\sqrt{1 + \left(\frac{r_\varphi}{r}\right)^2 + (r_z)^2}\,d\varphi' + \sin\left(\gamma - \frac{\pi}{2}\right)\rho\alpha dz - A_0,$$

where the initial effective surface area with no fluctuation is

$$A_0 = 2L\Gamma = 2L\left(r_0\varphi_0 + \sin\left(\gamma - \frac{\pi}{2}\right)\rho\alpha_0\right).$$

Taylor expanding the nonlinear terms and integrating over $\varphi'$ and $L$, we get

$$\delta A_0 = Lr_0\left[2d_1{}^2(1-s^2)\left(\varphi_0 + \frac{\sin(2s\varphi_0)}{2s}\right) - 4d_1{}^2\left(\tan\left(\gamma - \frac{\pi}{2}\right) + C\right)\cos^2(s\varphi_0)\right].$$

Substituting into the expression for $\delta F_{me}$, we arrive at the free energy variation caused by the fluctuation of the liquid meniscus for all elliptical grooves:

$$\delta F_{me} = 2d_1{}^2 L r_0 \left[ (\Delta p r_0 + \gamma_{LG})(1-s^2)\left(\varphi_0 + \frac{\sin(2s\varphi_0)}{2s}\right) \right.$$

$$- 2(\Delta p r_0 + \gamma_{LG})\left(\tan\left(\gamma - \frac{\pi}{2}\right) + C\right)$$

$$\left. + \Delta p r_0 \left(s\tan(s\varphi_0) - \tan\left(\gamma - \frac{\pi}{2}\right)\right)\cos^2(s\varphi_0) \right],$$

where $s^2 \in (-\infty, 1)$, $\alpha_0 \in (0, \pi/2)$, $\gamma \in (0, \pi)$, $\varphi_0 = \alpha_0 + \gamma - \pi/2$, and $\varphi_0 \in (-\pi/2, \pi)$.

### 2.4 Meniscus stability analysis

The stability of the meniscus with respect to the fluctuations is the condition $\delta F_{me} \geq 0$. Following the Young-Laplace equation, the first two terms in the square brackets are identically zero for both concave ($\Delta p > 0$, $r_0 < 0$, $\gamma < \pi/2$, and $\varphi_0 < 0$) and convex menisci ($\Delta p < 0$, $r_0 > 0$, $\gamma < \pi/2$, and $\varphi_0 < 0$). Hence, the condition $\delta F_{me} > 0$ reduces to

$$\Delta p \left[ s\tan(s\varphi_0) - \tan\left(\gamma - \frac{\pi}{2}\right) \right] > 0$$

for all $s$. To determine the ranges of $\varphi_0$ and $\gamma$, we analyze the function $\eta = \eta(s, \varphi_0, \gamma)$:

$$\eta(s, \varphi_0, \gamma) = s\tan(s\varphi_0) - \tan\left(\gamma - \frac{\pi}{2}\right). \tag{S5}$$

A simple analysis of **Eq. S5** yields the following conclusions (*S6*):

(1) The only stationary point of $\eta(s)$ is at $s = 0$, where $d^2\eta/ds^2 = 0$;
(2) In the absence of a singularity, $\eta$ is an increasing function as $s$ increases from 0 to 1;
(3) $\eta$ has a singularity at $s = \pi/(2\varphi_0)$. That is, it is singular for $\pi/2 < \varphi_0 < \pi$;
(4) For $s^2 = -\infty$, $\eta = +\infty$ for $\varphi_0 < 0$ and $\eta = -\infty$ for $\varphi_0 > 0$;
(5) The term $C$ that captures the effect of the groove geometry vanishes in the second term in the square brackets, indicating that the groove geometry is irrelevant to the free energy variation, and the criterion for the positive free energy variation ($\Delta p[s\tan(s\varphi_0) - \tan(\gamma - \pi/2)]$) applies to all groove geometries.

### 2.4.1. Positive Laplace pressures

For positive $\Delta p$, the stability condition requires that $\eta$ is positive, i.e., $\eta(0, \varphi_0, \gamma) > 0$ and $\eta(1, \varphi_0, \gamma) > 0$. This implies the following conditions:

(1) The liquid meniscus is concave, i.e., $r_0 < 0$, $\gamma < \pi/2$, and $-\pi/2 < \varphi_0 < 0$;

(2) $[\tan(\varphi_0) - \tan(\gamma - \pi/2) > 0] \cap [-\tan(\gamma - \pi/2) > 0]$ for $-\pi/2 < \varphi_0 < 0$;

(3) No singularity exists.

**2.4.2. Negative Laplace pressures**

For negative $\Delta p$, we have the following conditions:

(1) The liquid meniscus is convex, i.e., $r_0 > 0$, $\gamma > \pi/2$, and $0 < \varphi_0 < \pi/2$;

(2) $[\tan(\varphi_0) - \tan(\gamma - \pi/2) < 0] \cap [-\tan(\gamma - \pi/2) < 0]$ for $0 < \varphi_0 < \pi/2$;

(3) No singularity exists.

Note that no solution exists for these conditions, or no stable continuous gas layer can form for $\gamma > \pi/2$. This is true even for additional air inflated. The gas layer tends to collapse into discrete air pockets or completely escape.

The results of the fluctuation analysis are summarized as a phase diagram of the stable and unstable cases and shown in **Fig. 4b** in the main text. For consistency with our previous results for the quasi-2D grooves, the gas contact angle $\gamma$ is expressed in terms of $\theta_Y$, i.e., $\gamma = \pi - \theta_Y$ and $\varphi_0$ is expressed in terms of $\alpha_0$, i.e., $\varphi_0 = \alpha_0 + \theta_Y - \pi/2$.

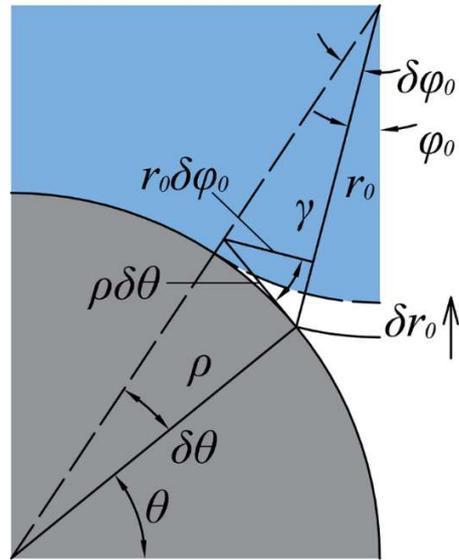

**Fig. S1.** The infinite small displacement $\delta r_0$ of the liquid meniscus that is part of the variational analysis. The solid line represents the initial state, and the dashed line represents that after displacement.

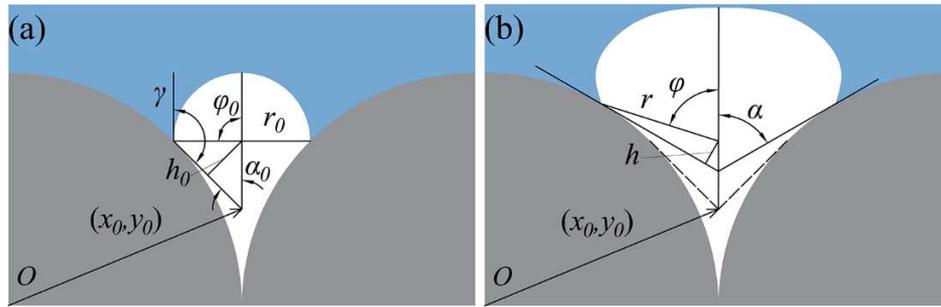

**Fig. S2.** The liquid-solid-gas cross-section showing the meniscus (a) initially and (b) subject to the azimuthal fluctuations considered in this study.

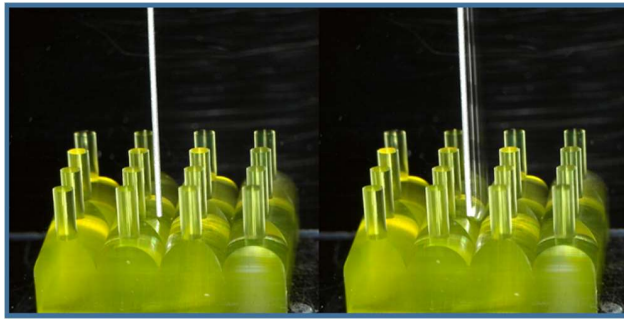

**Fig. S3.** The bubble escape process on the SHPOS with no hydrophobic coating with contact angle $\theta_Y = 73 \pm 1.1°$.

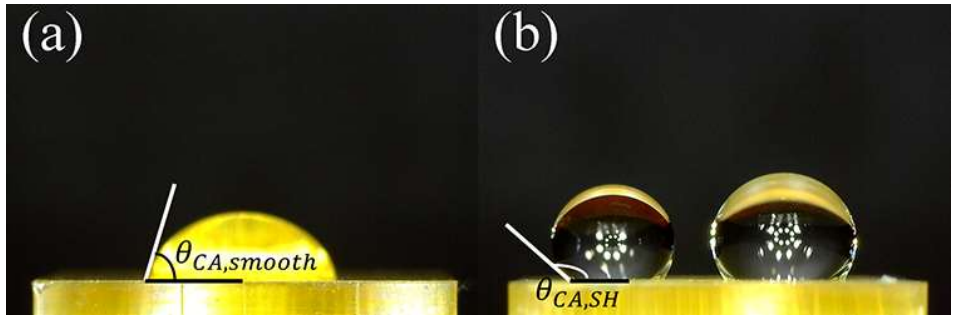

**Fig. S4.** (a) The liquid contact angle on the smooth surface, measured to be $73 \pm 1.1°$. (b) The liquid contact angle on the SHPOS coated with Glaco is $141 \pm 2.3°$.

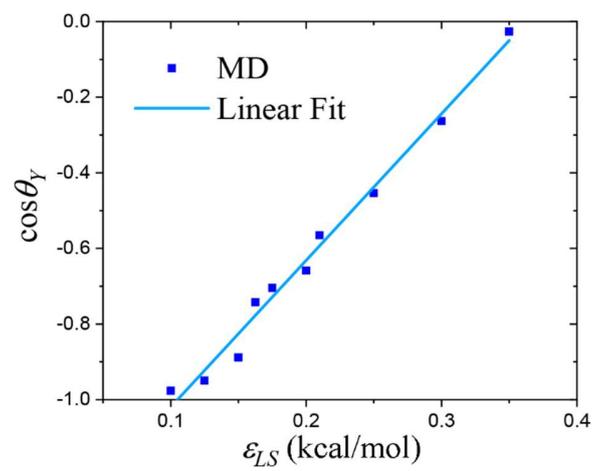

**Fig. S5.** The variation of the liquid contact angle $\cos\theta_Y$ plotted as a function LJ liquid-solid bond energy $\varepsilon_{LS}$. The blue line represents the linear fit.

**Table S1.** The Leonard-Jones parameters for solid, liquid, gas phases.

| Atoms | $\sigma_{ij}$ (Å) | $\varepsilon_{ij}$ (kcal/mol) |
|:---:|:---:|:---:|
| Liquid-Liquid (L-L) | 3.405 | 0.717 |
| Liquid-Gas (L-G) | 4 | 0.373 |
| Gas-Gas (G-G) | 4 | 0.191 |
| Liquid-Solid (L-S) | 3.405 | – |
| Gas-Solid (G-S) | 4 | 0.163 |

**Table. S2** The value of $\varepsilon_{LS}$ in the LJ potential and their corresponding liquid contact angles (with errors).

| $\varepsilon_{LS}$ (Kcal/mol) | $\theta_Y$ (°) | Err (°) |
|---|---|---|
| 0.1 | 167.70 | 0.50 |
| 0.125 | 161.70 | 1.15 |
| 0.15 | 152.68 | 1.50 |
| 0.175 | 134.78 | 4.00 |
| 0.2 | 131.20 | 1.50 |
| 0.21 | 124.42 | 2.80 |
| 0.25 | 117.00 | 2.50 |
| 0.3 | 105.26 | 2.70 |
| 0.35 | 91.52 | 2.25 |